\documentclass{article}

\usepackage{PRIMEarxiv}

\usepackage[utf8]{inputenc} 
\usepackage[T1]{fontenc}    
\usepackage{hyperref}       
\usepackage{url}            
\usepackage{booktabs}       
\usepackage{amsfonts}       
\usepackage{nicefrac}       
\usepackage{microtype}      
\usepackage{lipsum}
\usepackage{fancyhdr}       
\usepackage{graphicx}       
\graphicspath{{media/}}     

\title{Transfer functions of FXLMS-based Multi-channel Multi-tone Active Noise Equalizers
\thanks{\textit{\underline{Citation}}: 
\textbf{Transfer functions of FXLMS-based Multi-channel Multi-tone Active Noise Equalizer, Miguel Ferrer, Maria de Diego, Gema Piñero, Amin Hassani, Marc Moonen, Alberto Gonzalez, arXiv:submit/4381075 [eess.AS].}} 
}

\author{
  Miguel~Ferrer, Maria~de Diego, Gema~Pi\~{n}ero, Alberto~Gonzalez \\
  Institute of Telecommunications and Multimedia Applications (iTEAM),  \\
  Universitat Polit\'{e}cnica de Val\'{e}ncia (UPV) \\
  Valencia, Spain\\
  \texttt{\{mferrer, mdediego, gpinyero, agonzal\}@dcom.upv.es} \\
   \And
  Amin~Hassani, Marc~Moonen \\
  STADIUS center for dynamical systems, signal processing and data analytics \\
  KU Leuven \\
  Leuven, Belgium\\
  \texttt{\{amin.hassani, marc.moonen\}@esat.kuleuven.be} \\
}

\begin{document}
\maketitle

\begin{abstract}
Multi-channel Multi-tone Active Noise Equalizers can achieve different user-selected noise spectrum profiles even at different space positions. They can apply a different equalization factor at each noise frequency component and each control point. Theoretically, the value of the transfer function at the frequencies where the noise signal has energy is determined by the equalizer configuration. 
In this work, we show how to calculate these transfer functions with a double aim: to verify that at the frequencies of interest the values imposed by the equalizer settings are obtained, and to characterize the behavior of these transfer functions in the rest of the spectrum, as well as to get clues to predict the convergence behaviour of the algorithm. The information provided thanks to these transfer functions serves as a practical alternative to the cumbersome statistical analysis of convergence, whose results are often of no practical use. 
\end{abstract}


\section{Notation}  
\label{s:intro}
%
%
%
%

Throughout this work the following notation will be used: scalars and signals are written in italic lowercase letters, $a$, values of the transfer functions in italic uppercase, $A$, vectors in lowercase and bold, $\mathbf{a}$, while matrices are shown in uppercase and bold, $\mathbf{A}$. The symbols and operators used are: $i=\sqrt{-1}$ and $TZ\{\cdot\}$ for the Z-transform, so $F(z)$ is used to represent $TZ\{f(n)\} $ .

\section{Multi-channel Multi-tone Active Noise Equalizer with Spatially Distributed User-selected Profiles} \label{section1}
The Multi-channel Multi-tone Active Noise Equalizer (ANE) with Spatially Distributed User-selected Profiles was proposed in~\cite{pendiente}. In this section, we revise the basic mathematical formulation of this algorithm, as well as the two strategies that can iteratively manage it (common pseudo-error and multiple pseudo-error).

First of all, we consider a multi-channel adaptive equalizer composed by J actuators (J adaptive filters) and K error sensors (K signals to equalize) that weigh a multi-frequency noise signal by an arbitrary equalization factor $\beta_{lk}$, $\beta_{lk}\neq 1$, for each frequency $l$ at each sensor $k$. That is, considering that the noise signal in each $k$-th sensor is given by
\begin{equation}\label{dk}
    d_k(n)=\sum_{l=1}^L A_{kl}cos(\omega_l n + \phi_{kl}) \, ,
\end{equation}
where $L$ means the number of different frequencies of the noise signal. The goal of the algorithm is to achieve a residual noise signal in each  $k$-th sensor as
\begin{equation}\label{ek}
    e_k(n)=\sum_{l=1}^L \beta_{lk} A_{kl}cos(\omega_l n + \phi_{kl}).
\end{equation}
To this end, the algorithm disposes of $L$ reference signals, each of them defined as
\begin{equation}\label{xk}
    x_l(n)=A_{ref,l}cos(\omega_l n + \phi_{ref,l}).
\end{equation}

Based on notch adaptive filters as proposed by Kuo in \cite{ar:kuo93a,ar:kuo93,ar:kuo95}, an early proposal was presented in \cite{ar:ded03,ar:gon06b}, where the values of $\beta_{lk}$ was the same for all sensors at those particular frequencies ($\beta_{lk}=\beta_{l}$). In this proposal, the adaptive filtered-x LMS strategy \cite{ar:ell87,ar:bja95} was employed to adapt the filter coefficients by considering each complex coefficient as two real ones (one  fed  by  the  in-phase  component  of  the  reference  signal  and  the  other  by  its  quadrature  component). In particular, we adapt recursively (based on the widely known LMS algorithm \cite{bo:wid85}) the  coefficients as:
\begin{equation}
\mathbf{w}(n) = \mathbf{w}(n-1) - \mu \nabla_{\mathbf{w}}\{(\mathbf{e}'(n))^T\mathbf{e}'(n)\},
\label{Wlms}
\end{equation}

where $\mathbf{w}(n)$ is defined as the $2JL\times 1$ vector
\begin{equation}
\mathbf{w}(n) = 
[\mathbf{w}_{11}^T(n) \; \ldots \;\mathbf{w}_{1J}^T(n) \;\ldots \; \mathbf{w}_{L1}^T(n) \;\ldots \;\mathbf{w}_{LJ}^T(n) ]^T
\label{wvect}
\end{equation}
with $\mathbf{w}_{lj}(n)=[ w_{lj}(n) \; \hat{w}_{lj}(n) ]^{T}$, the in-phase and quadrature components of the filter coefficients at the actuator $j$ for the $l$-frequency. 
$\mu$ is the step size to assure stability \cite{bo:wid85}, and the $2JL\times 1$ vector $\nabla_{\mathbf{w}}$ represents the gradient operator:
\begin{equation}
\left[ \displaystyle{\frac{\partial}{\partial w_{11}}} \; \displaystyle{\frac{\partial}{\partial\hat{w}_{11}}}  \ldots  \displaystyle{\frac{\partial}{\partial w_{1J}}} \; \displaystyle{\frac{\partial}{\partial \hat{w}_{1J}}}  \ldots 
\displaystyle{\frac{\partial}{\partial w_{L1}}} \; \displaystyle{\frac{\partial}{\partial \hat{w}_{L1}}}  \ldots 
\displaystyle{\frac{\partial}{\partial w_{LJ}}} \; \displaystyle{\frac{\partial}{\partial \hat{w}_{LJ}}} \right]^T.
\label{gradiente}
\end{equation}
The pseudo-error vector $\mathbf{e'}(n)$ is a vector formed by the samples of the $K$ pseudo-error signals at time~$n$
\begin{equation}
    \mathbf{e'}(n)= [ e'_{1}(n)\;e'_{2}(n)\;\ldots\;e'_{K}(n)]^T \, ,
    \label{eq:vpseudoerror}
\end{equation}
where two strategies for the pseudo-error signals are considered, as it was introduced in \cite{ar:ded03}: 
\begin{itemize}
\item the common pseudo-error, which is defined as
\begin{equation}
e'_{k}[n]=e_k[n]+\sum_{l=1}^{L}\sum_{j=1}^J{\frac{1-\gamma_{lj}}{1-\beta_{lk}}\beta_{lk}
\widetilde{\mathbf{x}}^T_{ljk}[n]\mathbf{w}_{lj}[n]},
\label{pseudodcom}
\end{equation}
leading to the following update equation:
\begin{equation}
\mathbf{w}_{lj}[n]=\mathbf{w}_{lj}[n-1]-2\mu\sum_{k=1}^K\frac{1-\gamma_{lj}}{1-\beta_{lk}}
\widetilde{\mathbf{x}}_{ljk}[n]e'_{k}[n]\}, \label{Wlmspsecom}
\end{equation}
\item the multiple pseudo-error, defined as
\begin{equation}
e'_{lk}(n)=e_k(n)+\sum_{j=1}^J{\frac{1-\gamma_{lj}}{1-\beta_{lk}}\beta_{lk}
\widetilde{\mathbf{x}}^T_{ljk}(n)\mathbf{w}_{lj}(n)},
\label{pseudodmul}
\end{equation}
leading to the following update equation:
\begin{equation}
\mathbf{w}_{lj}(n)=\mathbf{w}_{lj}(n-1)-2\mu\sum_{k=1}^K\frac{1-\gamma_{lj}}{1-\beta_{lk}}
\widetilde{\mathbf{x}}_{ljk}(n)e'_{lk}(n). \label{Wlmspsemul}
\end{equation}
\end{itemize}

We have considered that the $j$-th filter outputs at the $l$-th frequency are weighted by $(1-\gamma_{lj})$ in order to adjust the signal dynamic range in practice.

Furthermore, we define the following signals:
\begin{equation}
\widetilde{\mathbf{x}}_{ljk}(n)=\left[\begin{array}{c}

A_{ljk} A_{\mathrm{ref,l}} \cos(\omega_l n+\phi_{\mathrm{ref,l}}+\phi_{ljk}) \\
A_{ljk}A_{\mathrm{ref,l}} \sin(\omega_l n+\phi_{\mathrm{ref,l}}+\phi_{ljk}) ,
\label{eq:vector_xjk}
\end{array}\right]
\end{equation}
where $A_{ljk}$ and $\phi_{ljk}$ are the magnitude and phase of an estimation of the transfer function of the acoustic path $C_{jk}$ that links the $j$-th actuator and the $k$-th error sensor evaluated at the $l$-th frequency.

Finally, we have
\begin{equation}
e_{k}(n)=d_k(n)+\sum_{l=1}^{L}\sum_{j=1}^J(1-\gamma_{lj})\mathbf{x}^T_{ljk}(n)\mathbf{w}_{lj}(n),
\label{errormf}
\end{equation}
 where $\mathbf{x}_{ljk}(n)$ is calculated as $\widetilde{\mathbf{x}}_{ljk}(n)$ but using the actual acoustic paths instead of their estimations or measurements. In case of perfect acoustic paths identification, then  $\widetilde{\mathbf{x}}_{ljk}(n)=\mathbf{x}_{ljk}(n)$. 
Further description of this algorithm can be found in~\cite{pendiente}. In this report, we introduce how to calculate the transfer functions of these equalizers considering both strategies in order to check the equalizer performance according to the design specifications (i.e., weighting the $l$ frequency noise signal at sensor $k$ by the weighting factor $\beta_{lk}$) and to obtain information of interest about the behavior of these equalizers, such as its frequency selectivity. 
\begin{figure}
\begin{center}
\includegraphics[width=0.55\textwidth]{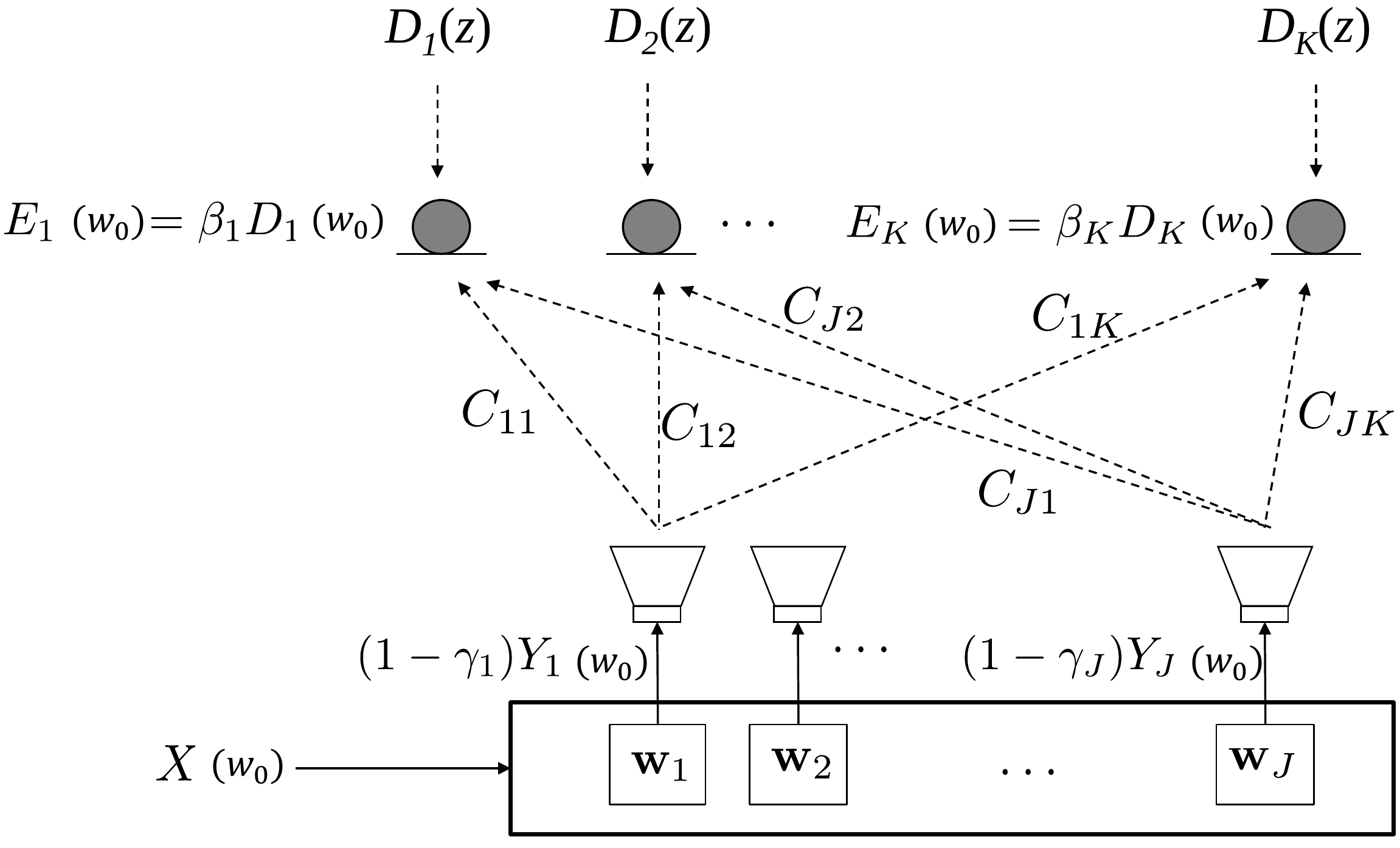}
\end{center}
\caption{Multi-channel system for the active control of a single-frequency noise with different equalization factors, $\beta_k$, at each control point.}\label{f:figura1bis}
\end{figure}

\section{Numerical calculation of the Transfer Functions} 
\label{s:anexo-ft}
The transfer function for a single sensor of the multi-channel system is defined as 
\begin{equation}
H_k(z)=\frac{E_k(z)}{D_k(z)}, \quad 1\leq k\leq K.
\label{eq:Hft}
\end{equation}
Therefore, there exist a different transfer function for each error sensor. Unfortunately, these transfer functions only present a closed expression for a single-channel system, i.e. \cite{ar:gon06b}, due to the dependence between the transfer function of each sensor within a generic multi-channel system and therefore they have to be numerically evaluated. 

\subsection{Multi-channel single-frequency active equalizer}
First of all, we consider a single frequency ANE. From a theoretical point of view, the absolute value of $H_k(z)$ evaluated at the noise frequency, $z=e^{i\omega_0}$, should be equal to the equalization gain $\beta_k$, as it is illustrated by Figure~\ref{f:figura1bis}. However, the calculation of the transfer function over the whole frequency band can give deeper knowledge on the behaviour of the ANE at those frequencies that are not being controlled because they are not part of the single frequency reference signal.

For sake of simplicity, we assume that the reference signal is a pure tone of unit amplitude, angular frequency $\omega_{0}$ and initial phase equal to $0$, such that: 
\begin{equation}
\begin{array}{c}
x(n)=\cos(\omega_{0}n)=\frac{1}{2}(e^{i\omega_{0}n}+e^{-i\omega_{0}n}), \\ \widehat{x}(n)=\sin(\omega_{0}n)=\frac{1}{2i}(e^{i\omega_{0}n}-e^{-i\omega_{0}n}).
\end{array}
\label{ececuft8}
\end{equation}

Therefore, the filter output signals $y_j(n)$ before they are weighted by its output factor ($1-\gamma_{j}$) (see Figure~\ref{f:figura1bis}) are calculated as:
\begin{equation}\begin{array}{l}
y_j(n)=w_j(n)x(n)+\widehat{w}_j(n)\widehat{x}(n)=
\frac{1}{2}\{w_j(n)(e^{i\omega_{0}n}+e^{-i\omega_{o}n})-i\widehat{w}_j(n)(e^{i\omega_{0}n}-e^{-i\omega_{0}n})\}.
\end{array}
\label{eq:ececuft9}
\end{equation}

The updating equation of the filter coefficients, given in (\ref{Wlmspsecom}) or (\ref{Wlmspsemul}) (they are equivalent for a single frequency) as a whole vector, can be particularized for a single frequency and expressed for each coefficient in the Z-transform domain as
\begin{eqnarray}
z W_j(z) & = & W_j(z)-2\mu\sum_{m=1}^K\frac{1-\gamma_j}{1-\beta_m}\mathrm{TZ}\{\widetilde{x}_{jm}(n)e'_m(n)\}, \label{eq:wz1} \\
z \widehat{W}_j(z) & = & \widehat{W}_j(z)-2\mu\sum_{m=1}^K\frac{1-\gamma_j}{1-\beta_m}\mathrm{TZ}\{\widetilde{\widehat{x}}_{jm}(n)e'_m(n)\}.
\label{eq:wz2}
\end{eqnarray}
Notice that the sum uses index $m$ to refer to the recording position whereas we use index $k$ to refer to the particular location where the transfer function $H_k(z)$ in (\ref{eq:Hft}) will be computed. Terms $\widetilde{x}_{jm}$ and $\widetilde{\widehat{x}}_{jm}$ are the in-phase and quadrature components of the reference signal filtered by the estimate of the corresponding secondary path $\widetilde{c}_{jm}$ at frequency $\omega_{0}$. $\widetilde{x}_{jm}$ and $\widetilde{\widehat{x}}_{jm}$ are also the elements of vector $\widetilde{\mathbf{x}}_{jm}$ that was defined in (\ref{eq:vector_xjk}). Considering the secondary path as $\widetilde{c}_{jm} = \widetilde{A}_{jm} e^{i\widetilde{\phi}_{jm}}$, then
\begin{eqnarray}
\mathrm{TZ}\{\widetilde{x}_{jm}(n)e'_m(n)\} & = & \frac{\widetilde{A}_{jm}}{2}[E'_m(ze^{-i\omega_{0}-\widetilde{\phi}_{jm}})+E'_m(ze^{i\omega_{0}+\widetilde{\phi}_{jm}})], \label{eq:ececuft1} \\ 
\mathrm{TZ}\{\widehat{\widetilde{x}}_{jm}(n)e'_m(n)\} & = & \frac{\widetilde{A}_{jm}}{2i}[E'_m(ze^{-i\omega_{0}-\widetilde{\phi}_{jm}})-E'_m(ze^{i\omega_{0}+\widetilde{\phi}_{jm}})].
\label{eq:ececuft2}
\end{eqnarray}

Substituting (\ref{eq:ececuft1}) and (\ref{eq:ececuft2}) in (\ref{eq:wz1}) and (\ref{eq:wz2}) respectively, we obtain:
\begin{eqnarray}
W_j(z) & = & \frac{-\mu}{z-1}\sum_{m=1}^K\widetilde{A}_{jm}\frac{1-\gamma_j}{1-\beta_m}[E'_m(ze^{-i\omega_{0}-\widetilde{\phi}_{jm}})+E'_m(ze^{i\omega_{0}+\widetilde{\phi}_{im}})], \label{eq:WecuI1} \\
\widehat{W}_j(z) & = & \frac{-\mu}{i(z-1)}\sum_{m=1}^K\widetilde{A}_{jm}\frac{1-\gamma_j}{1-\beta_m}[E'_m(ze^{-i\omega_{0}-\widetilde{\phi}_{jm}})-E'_m(ze^{i\omega_{0}+\widetilde{\phi}_{jm}})].
\label{eq:WecuI2}
\end{eqnarray}

On the other hand, the output of the $j$-th adaptive filter given by (\ref{eq:ececuft9}) is expressed in the Z-transform domain as
\begin{equation}
Y_j(z)=\frac{1}{2}\{W_j(z e^{-i\omega_{0}})+W_j(z e^{i\omega_{0}})-i
\widehat{W}_j(z e^{-i\omega_{0}})+i\widehat{W}_j(z e^{i\omega_{0}})
\}. \label{ececuft12}
\end{equation}

Substituting (\ref{eq:WecuI1}) and (\ref{eq:WecuI2}) in (\ref{ececuft12}) and arranging the result, we 
get
\begin{equation}
{Y}_j(z)=-2\mu\sum_{m=1}^K\widetilde{A}_{jm}\frac{1-\gamma_j}{1-\beta_m}\frac{z \cos(\omega_{0}-\widetilde{\phi}_{jm})-\cos(\widetilde{\phi}_{jm})}{z^{2}-2z \cos(\omega_{0})+1}E'_m(z),
\label{ececuft17}
\end{equation}
that can be expressed in compact form as 
\begin{equation}
{Y}_j(z)=\sum_{m=1}^K G_{jm}(z)E'_m(z),
\label{eq:Yjz}
\end{equation}
where $G_{jm}(z)$ is an auxiliary function defined as:
\begin{equation}
G_{jm}(z)=-2\mu\widetilde{A}_{jm}\frac{1-\gamma_j}{1-\beta_m}\frac{z \cos(\omega_{0}-\widetilde{\phi}_{jm})-\cos(\widetilde{\phi}_{jm})}{z^{2}-2z \cos(\omega_{0})+1}.
\label{eq:Gjm}
\end{equation}

\begin{figure}
\begin{center}
 \includegraphics[width=0.48\textwidth]{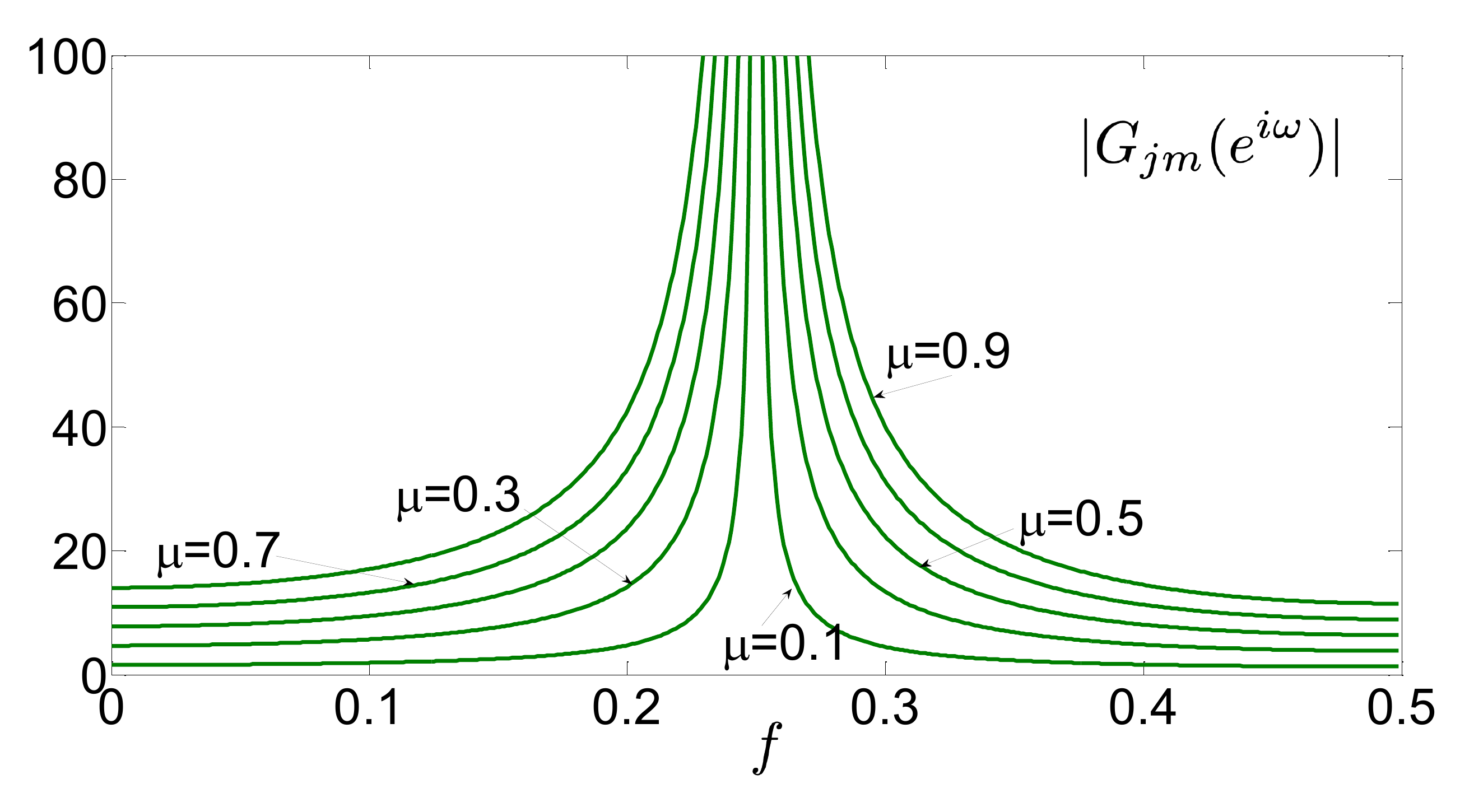} \hspace{2ex}
  \includegraphics[width=0.45\textwidth]{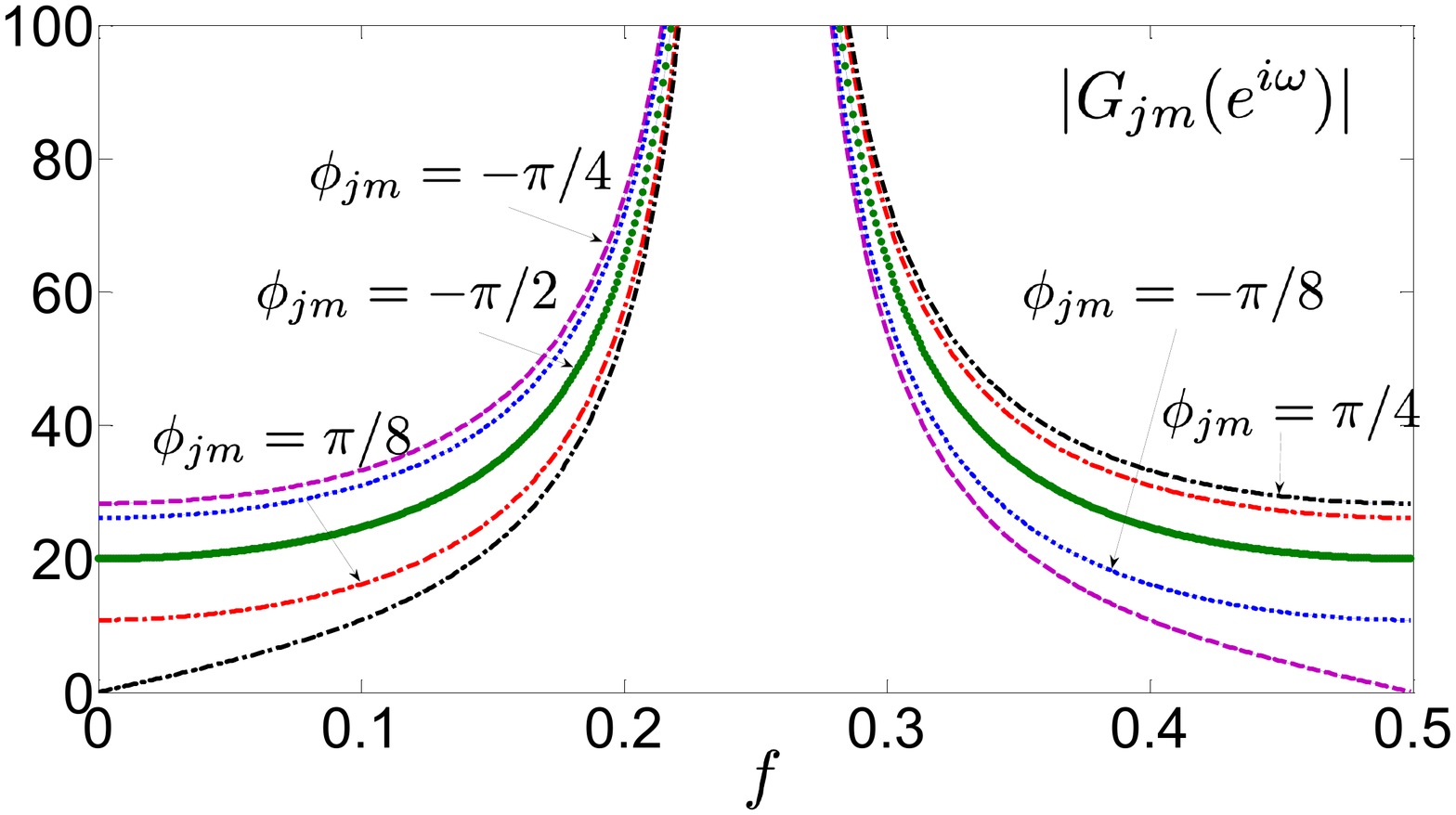}
\end{center}
\caption{Performance of the frequency response of the auxiliary functions $G_{jm}(e^{i\omega})$ with $\mu$ (left plot), and $\phi_{jm}$ (right plot). We have considered: $\omega_0=\pi/4$, $\widetilde{A}_{jm}=1$, $\gamma_j=0$ and $\beta_m=0.5$.}\label{f:figure3}
\end{figure}

It can be observed that the auxiliary functions $G_{jm}(z)$ have a pole at $\omega=\omega_{0}$. Regarding the frequency response $G_{jm}(e^{i\omega})$ around $\omega=\omega_{0}$, the width of the function grows with $\left|\mu\widetilde{A}_{jm} \frac{1-\gamma_j}{1-\beta_m}\right|$. We can fix particular values to: $\omega_0$, $\widetilde{A}_{jm}$, $\gamma_j$ and $\beta_m$; to study the influence of different parameters on the auxiliary functions defined by (\ref{eq:Gjm}), i.e. $\omega_0=\pi/4$, $\widetilde{A}_{jm}=1$, $\gamma_j=0$ and $\beta_m=0.5$. As it can be appreciated in the left plot of Figure~\ref{f:figure3}, the larger the convergence step $\mu$, the wider the peak due to the pole. Therefore, the convergence of the adaptive algorithm would be faster as $\mu$ grows, but due to the increasing of the pole width, it would also be less selective in frequency. On the other hand, the value of $\widetilde{\phi}_{jm}$ rules the lack of symmetry of $G_{jm}(e^{i\omega})$ as the right plot of Figure~\ref{f:figure3} shows. Notice that the general behavior of functions $G_{jm}(z)$ is independent of the particular frequency $\omega_0$ or weighting parameter $\beta_m$ chosen, it depends only on the value of $\mu$ and the secondary paths $\widetilde{c}_{jm} = \widetilde{A}_{jm} e^{i\widetilde{\phi}_{jm}}$. It should be also noted that the frequency response of the transfer functions $H_k(z)$ in (\ref{eq:Hft}) will depend on the combination of the different functions $G_{jm}(z)$, as it shows the expression of ${Y}_j(z)$ in (\ref{eq:Yjz}).

At this point, we substitute the pseudo-error term $E'_m(z)$ in (\ref{eq:Yjz}) by their value given by the Z-transform of $e'_{m}(n)$ in (\ref{Wlmspsecom}), considering $k=m$ and a single frequency case:
\begin{equation}
Y_j(z)=\sum_{m=1}^K \{G_{jm}(z)E_m(z)+G_{jm}(z)\sum_{r=1}^J
\frac{1-\gamma_r}{1-\beta_m}\beta_m Y_r(z)\widetilde{C}_{rm}(z)\},
 \label{ececuftm7b}
\end{equation}
and gather in the left side of the equation the components multiplied by $Y_j(z)$, leading to:
\begin{eqnarray}
 Y_j(z) (1-\sum_{m=1}^K \frac{1-\gamma_j}{1-\beta_m}\beta_m G_{jm}(z) \widetilde{C}_{jm}(z)) = 
 \sum_{m=1}^K (G_{jm}(z)E_m(z) \nonumber \\
 + G_{jm}(z) \sum_{r=1, r\neq j}^J \frac{1-\gamma_r}{1-\beta_m} \beta_m Y_r(z) \widetilde{C}_{rm}(z)).
\label{ezkex11}
\end{eqnarray}

On the other hand, there exists a direct relation between the error signal $E_m(z)$ and the contribution of the noise at the $m$th position $D_m(z)$ given by: 
\begin{equation}
E_m(z)=D_m(z)+\sum_{j=1}^J{(1-\gamma_j)Y_j(z) \widetilde{C}_{jm}(z)},
\label{ezkex12}
\end{equation}
where we have considered again estimated secondary paths and $k=m$.

Returning to the expression of $H_k(z)=\frac{E_k(z)}{D_k(z)}$ for $k=1,\ldots,K$, the $K$ unknown terms are $E_k(z)$, which in turn depend on the $J$ unknown terms $Y_j(z)$, and the available $K+J$ equations that involve all the unknowns are given by (\ref{ezkex11}) and (\ref{ezkex12}). However, we can state a direct relation with the transfer function of interest $H_k(z)$ dividing by $E_k(z)$ the $K$ equations in (\ref{ezkex12}):
\begin{equation}
\frac{E_m(z)}{E_k(z)}=\frac{D_m(z)}{E_k(z)}+\sum_{j=1}^J{(1-\gamma_j)\frac{Y_j(z)}{E_k(z)}C_{jm}(z)},
\label{sistk}
\end{equation}
 and the $J$ equations in (\ref{ezkex11}):
\begin{equation}\begin{array}{l}
\displaystyle{\frac{Y_j(z)}{E_k(z)}\left[1-\sum_{m=1}^K\frac{1-\gamma_j}{1-\beta_m}\beta_m G_{jm}(z)\widetilde{C}_{jm}(z)\right]}=\\
\displaystyle{\sum_{m=1}^K \{G_{jm}(z) \frac{E_m(z)}{E_k(z)} + G_{jm}(z) \sum_{r=1,r\neq j}^J \frac{1-\gamma_r}{1-\beta_m} \beta_m \frac{Y_r(z)}{E_k(z)} \widetilde{C}_{rm}(z)\}}.
\end{array}
\label{ececuftm7d}
\end{equation}

Notice that the unknown transfer function $H_k(z)$ is implicitly included in (\ref{sistk}) since:
\begin{equation}
\frac{D_m(z)}{E_k(z)}=\frac{D_m(z)}{D_k(z)}\frac{D_k(z)}{E_k(z)}=\frac{P_m(z)}{P_k(z)}\frac{1}{H_k(z)},
\label{1H}
\end{equation}
whereas the other $K+J-1$ unknowns of are
$\displaystyle{\frac{E_m(z)}{E_k(z)}}$ for $m\neq k$, and
$\displaystyle{\frac{Y_j(z)}{E_k(z)}}$ for $1\leq j\leq J$.

The equation system given by (\ref{sistk})-(\ref{ececuftm7d}) can be expressed in matrix form as
\begin{equation}
\begin{array}{l l l}
\left[ \begin{array}{c c } \mathbf{A}_{K\times K} &\mathbf{ B}_{K\times J} \\
\mathbf{D}_{J\times K} & \mathbf{E}_{J\times J}
 \end{array} \right] & \left[\begin{array}{c}
\mathbf{f}_{K\times 1} \\
\mathbf{g}_{J\times 1}
 \end{array}\right] &  =\left[\begin{array}{c}
\mathbf{u}_{K\times 1} \\
\mathbf{v}_{J\times 1}
 \end{array}\right],
 \end{array}
\label{comp}
\end{equation}

where the matrices and vectors appearing in (\ref{comp}) would depend only on the equalization weights ($\beta_k$), output weights ($\gamma_j$), and the values of the transfer functions of the acoustic paths and their estimates (which also appear in the auxiliary functions $G_{jk}(z)$). Thus, for the transfer function $H_1(z)$ of a generic system with $J$ actuators and $K$ sensors, we would have that the elements of row $f$ and column $c$ of the matrix $\mathbf{A}_{K\times K}$ of size $K\times K$ could be calculated according to:

\begin{equation}\label{eleA}
    \begin{array}{l}
      a_{f1}=P_f(z)/P_1(z) \\
      a_{ff}=-1, \qquad\qquad f\neq 1 \\
      a_{fc}=0, \qquad\qquad \rm{other cases}. \\
    \end{array}
\end{equation}

The elements of the matrix $\mathbf{B}_{K\times J}$ are obtained by mean of:
\begin{equation}\label{eleB}
      b_{fc}=(1-\gamma_c)C_{cf}(z),
\end{equation}
meanwhile, the elements of the matrix $\mathbf{D_{J\times K}}$ are given by:
\begin{equation}\label{eleD}
\begin{array}{l}
  d_{fc}=0, \qquad\qquad  c = 1  \\
    d_{fc}=-G_{fc}(z), \qquad\qquad  c\neq 1,
\end{array}
\end{equation}

and those of the matrix $\mathbf{E}_{J\times J}$ by:
\begin{equation}\label{eleE}
\begin{array}{l}
  \displaystyle{e_{fc}=1-(1-\gamma_c)\sum_{k=1}^K \frac{\beta_k}{1-\beta_k}G_{fk}(z)\widetilde{C}_{ck}(z)},\qquad  f = c  \\
   \displaystyle{e_{fc}=-(1-\gamma_c)\sum_{k=1}^K \frac{\beta_k}{1-\beta_k}G_{fk}(z)\widetilde{C}_{ck}(z)},\qquad  f\neq c.
\end{array}
\end{equation}

The vectors containing the unknowns will be:
\begin{equation}
\mathbf{f}_{K\times 1}^T=[f_1\;f_2\;\ldots\;
f_K]=\left[\frac{1}{H_1(z)}\;\frac{D_2(z)}{E_1(z)}\;\ldots\;\frac{D_K(z)}{E_1(z)}\right],
\label{vF}
\end{equation}
\begin{equation}
\mathbf{g}_{J\times 1}^T=[g_1\;g_2\;\ldots\;
g_J]=\left[\frac{Y_1(z)}{E_1(z)}\;\frac{Y_2(z)}{E_1(z)}\;\ldots\;\frac{Y_J(z)}{E_1(z)}\right],
\label{vg}
\end{equation}

while we will have for the $K$ elements of $\mathbf{u}_{K\times 1}$ and the $J$ elements of $\mathbf{v}_{J\times 1}$ the following values:
\begin{equation}\label{eleU}
\begin{array}{l}
  u_{1}=1, \\
  u_{f}=0 \qquad  (f\neq 1),
\end{array}
\end{equation}

and
\begin{equation}\label{eleV}
\begin{array}{l}
  v_{f}=G_{f1}(z),\\
\end{array}
\end{equation}

respectively.

Considering the matrices of (\ref{comp}) and the values of their components, it is possible to solve the system (particularized for given values of $z$) as:
\begin{equation}
\begin{array}{l l l}
 \left[\begin{array}{c}
\mathbf{f}_{K\times 1} \\
\mathbf{g}_{J\times 1}
 \end{array}\right] &  =\left[ \begin{array}{c c } \mathbf{A}_{K\times K} & \mathbf{B}_{K\times J} \\
\mathbf{D}_{J\times K} & \mathbf{E}_{J\times J}
 \end{array} \right]^{-1} &\left[\begin{array}{c}
\mathbf{u}_{K\times 1} \\
\mathbf{v}_{J\times 1}
 \end{array}\right],
 \end{array}
\label{solcomp}
\end{equation}
where we are only concerned in the first element of the solution of (\ref{solcomp}), which provides: $H_1(z)=1/f_{1}(z)$.

The above procedure can be repeated to obtain any generic transfer function of the form $H_m(z)=E_m(z)/D_m(z)$ by
re-stating the system (\ref{comp}), and re-defining matrices and vectors. We can also start from the solution proposed for $H_1(z)$, exchanging the subscript 1 for $n$, in the functions $C_{jk}(z)$,
$G_{jk}(z)$, $G_{jk}(z)$ or $P_k(z)$, in the equalization profiles $\beta $, and in the output weights $\gamma_j$. This way, we can take profit of the matrices and vectors used to get $H_1(z)$. 
For a single-channel system (only one loudspeaker and one sensor), it is straightforward to obtain its single transfer function as:
\begin{equation}
H_1(z) = \displaystyle{\frac{1-\frac{1-\gamma_1}{1-\beta_1}\beta_1G_{11}(z)\tilde{C}_{11}(z)}{1-\frac{1-\gamma_1}{1-\beta_1}G_{11}(z)C_{11}(z)+\frac{1-\gamma_1}{1-\beta_1}\beta_1G_{11}(z)[C_{11}(z)-\tilde{C}_{11}(z)]}}.
\label{solmono}
\end{equation}

\begin{figure}
\begin{center}
\includegraphics[width=0.65\textwidth]{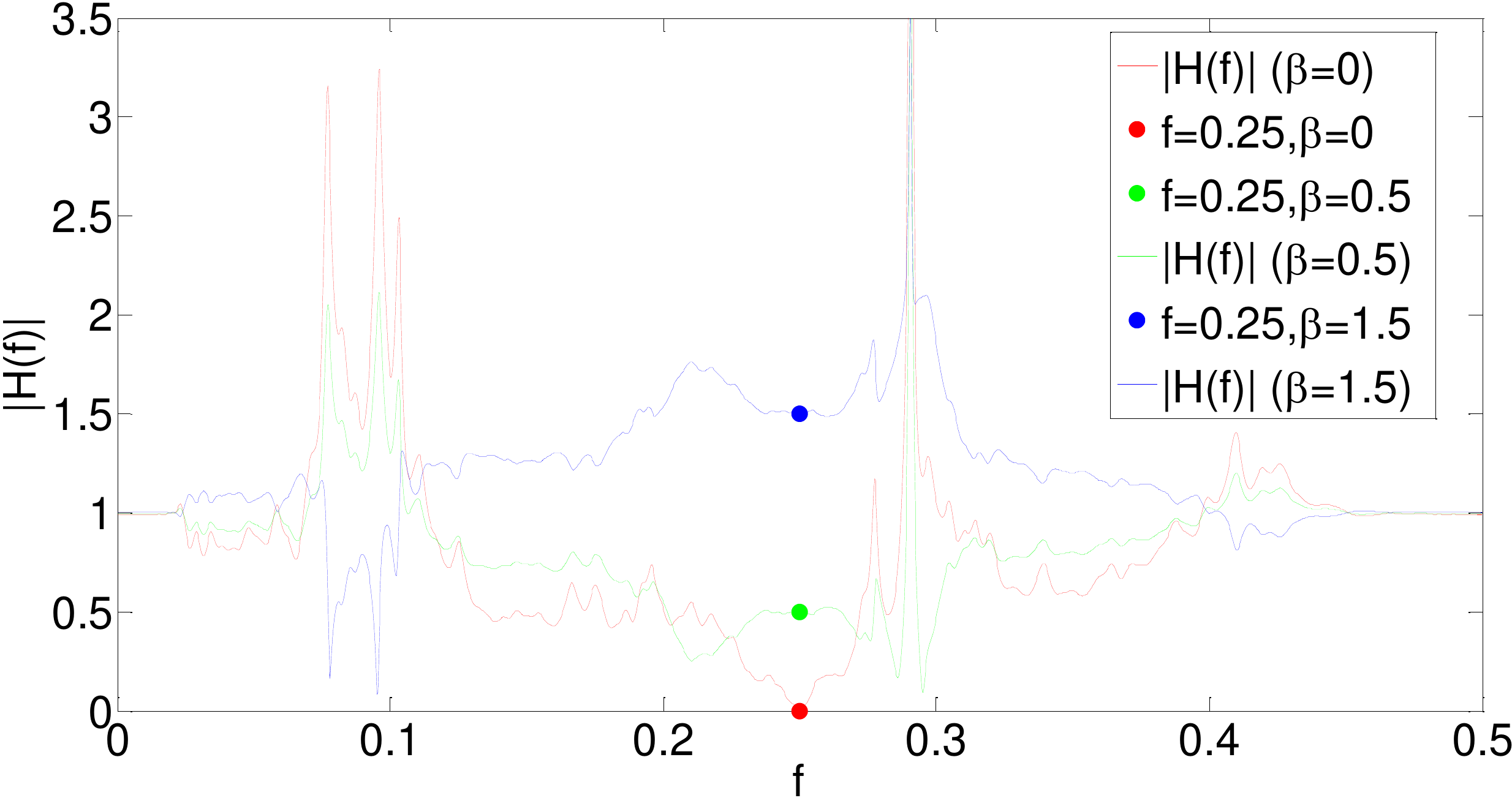}     
\end{center}
\caption{Magnitude of the frequency response for a single-channel ANE, with $\beta\in\left\{0, 0.5, 1.5\right\}$, at discrete frequency 0.25.}\label{f:figure6}
\end{figure}

\begin{figure}
\begin{center}
\begin{tabular}{cc }
\includegraphics[width=8cm]{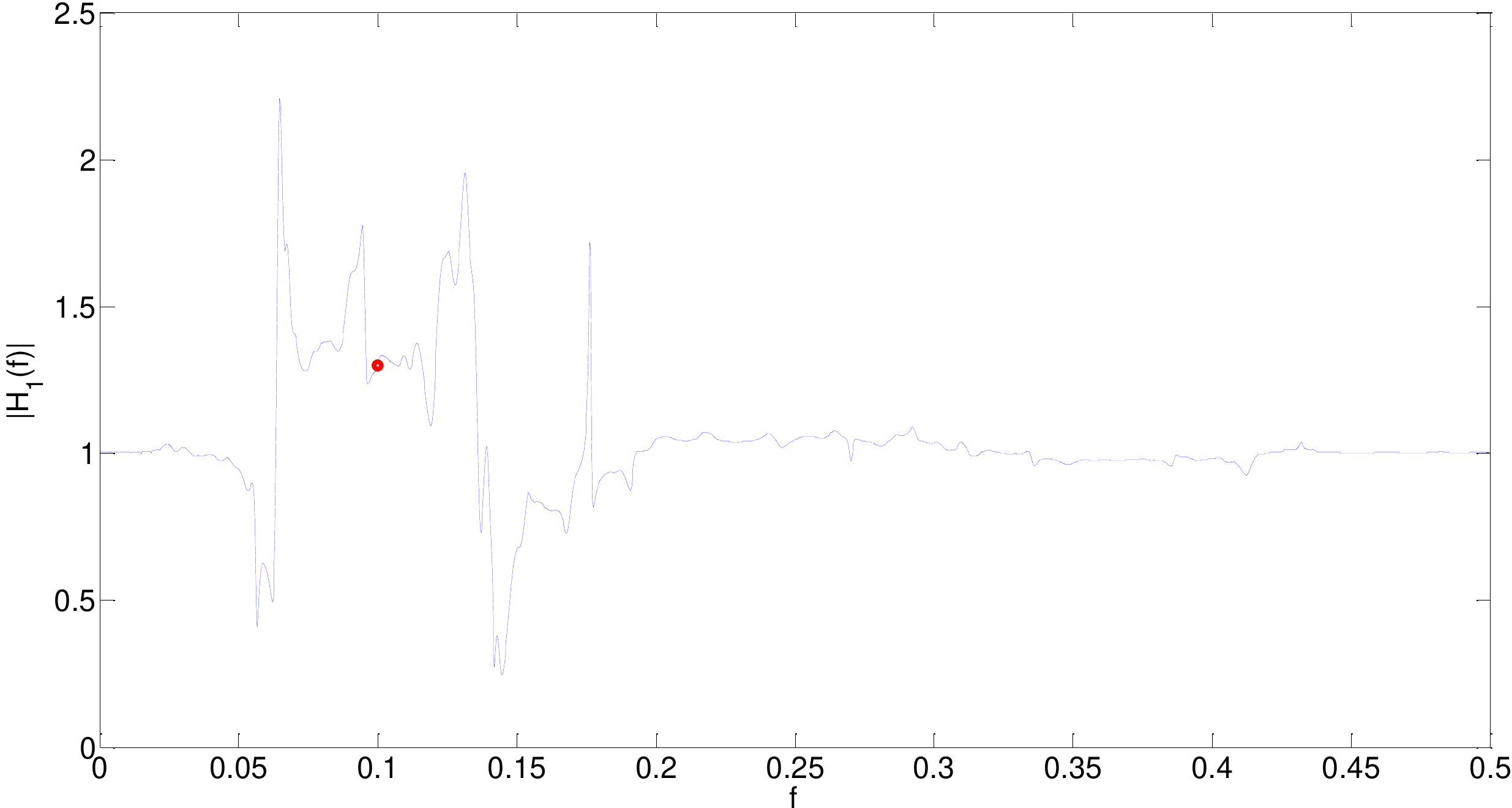} &
\includegraphics[width=8cm]{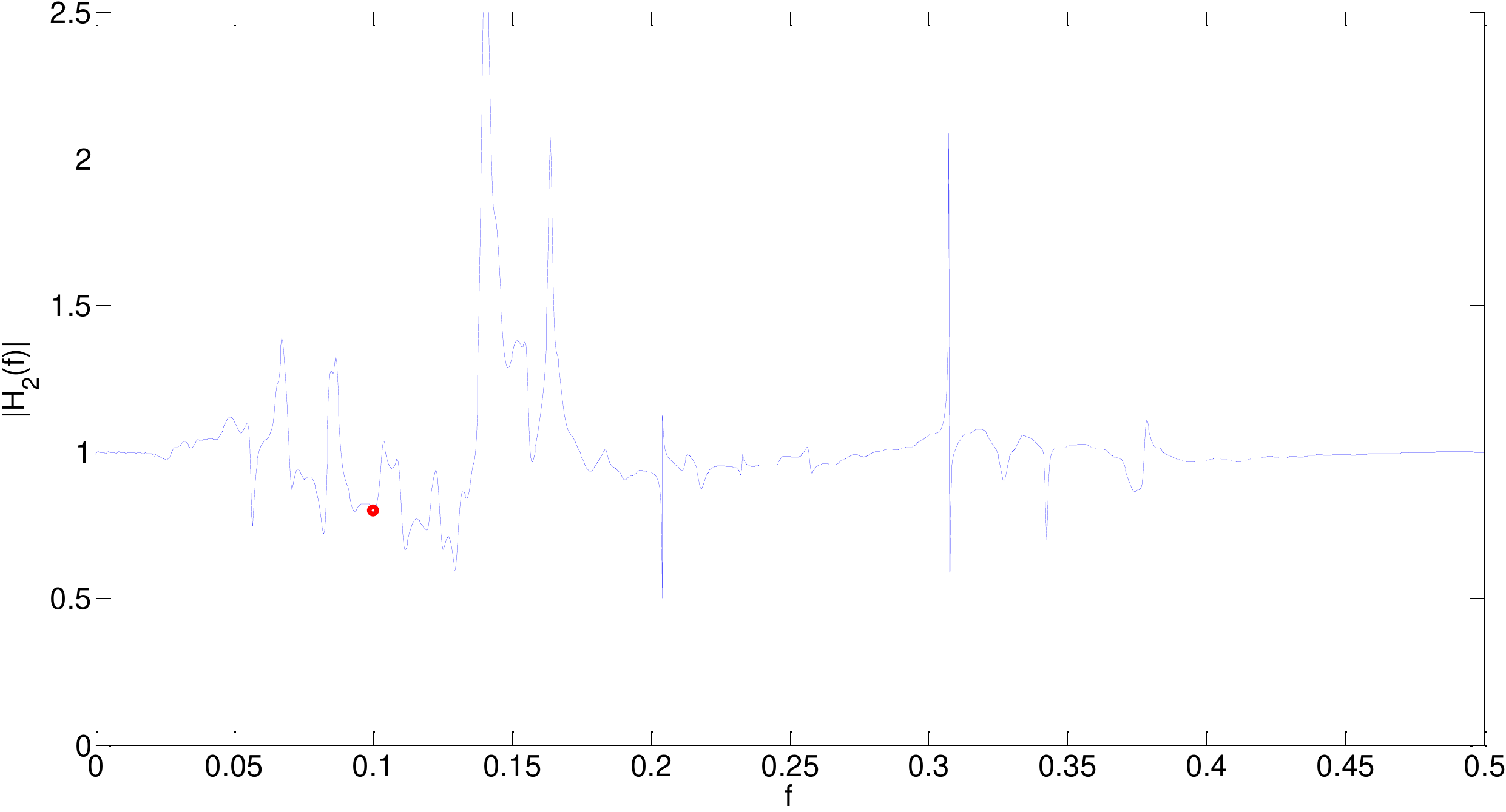}\\
Sensor 1:$\beta_1=1.3$. &  Sensor 2:$\beta_2=0.8$ \\
\includegraphics[width=8cm]{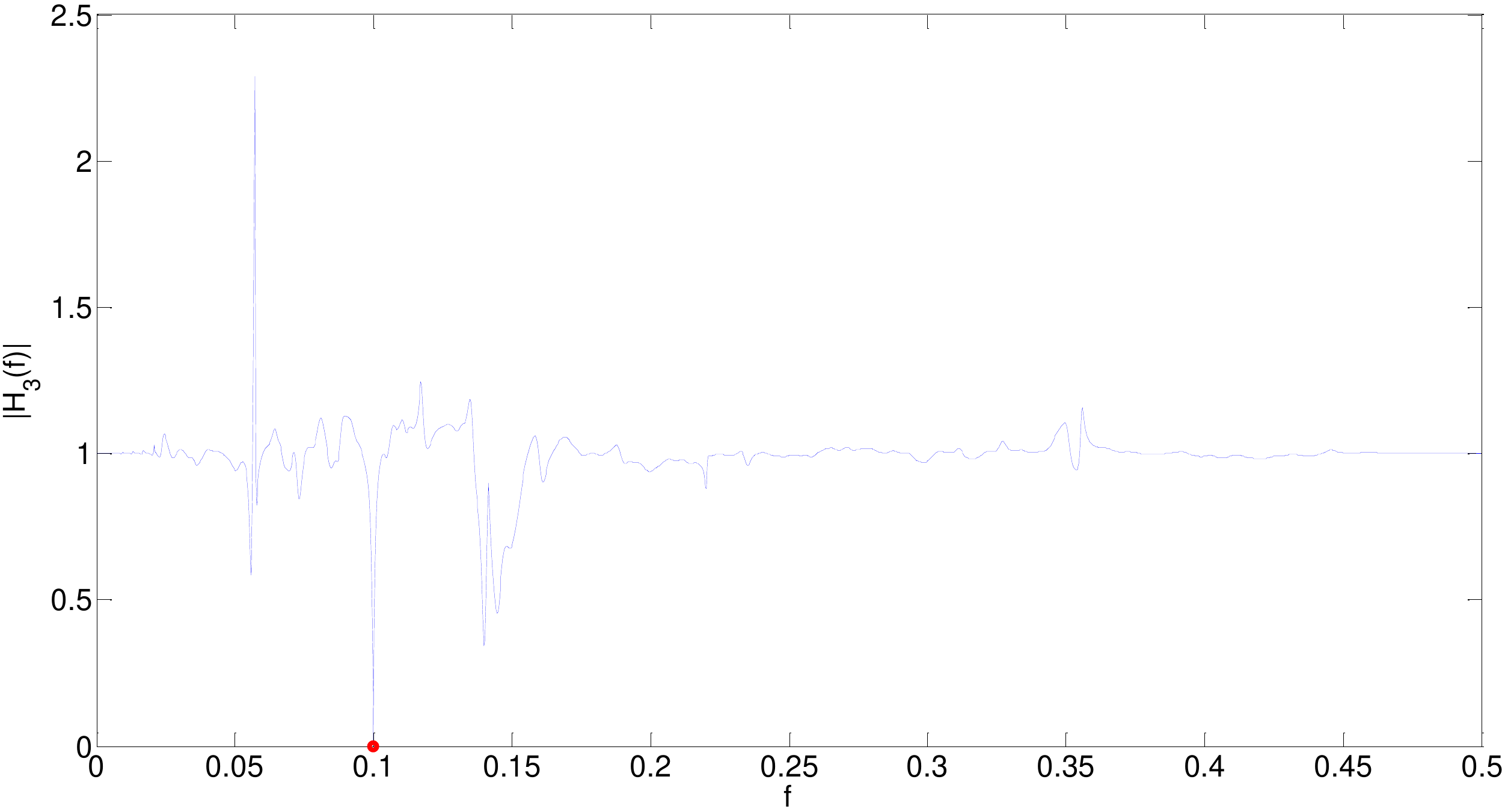} &
\includegraphics[width=8cm]{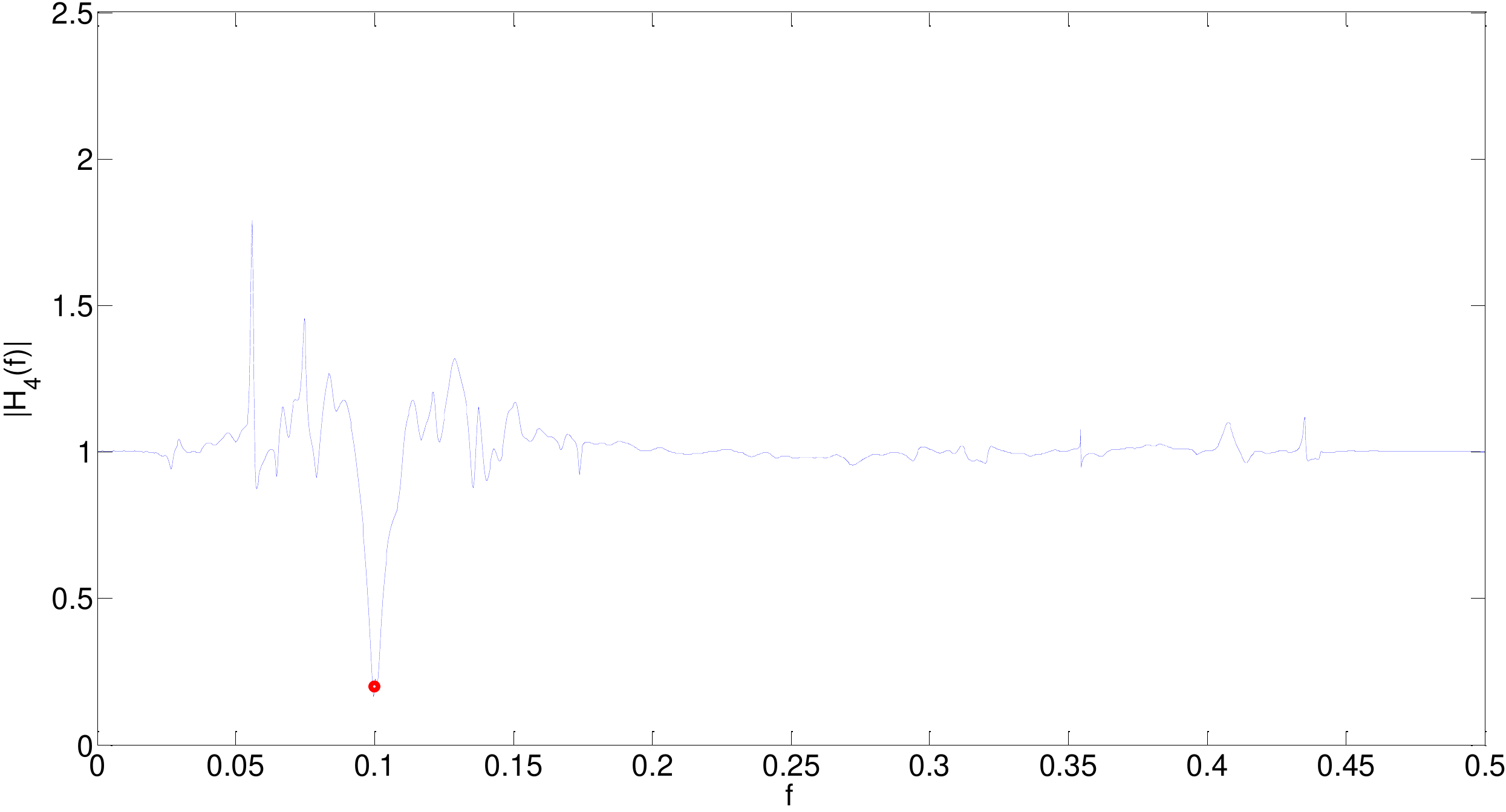}  \\
Sensor 3:$\beta_3=0$. &  Sensor 4:$\beta_4=0.2$
\end{tabular}
\end{center}
\caption{Magnitude of the frequency response for a multi-channel ANE, with 4 sensors and 4 loudspeakers, and different equalization profiles in each sensor (marked by a red spot).}\label{f:figure7}
\end{figure}

Figure \ref{f:figure6} shows the magnitude of the frequency response of a single-channel system. It is evaluated for different equalization weights at the discrete frequency $f=0.25$. Figure~\ref{f:figure7} shows the magnitude of the different frequency response of a multi-channel system with four sensors and four loudspeakers, where in each sensor a different equalization weight is applied at discrete frequency $f=0.1$; $\gamma_j=0$, $\forall j$, in both cases. It is found that  the desired transfer function value is obtained for the control frequency in all cases, while this response depends on the frequency responses of the acoustic paths, among other factors, for the rest of the the frequencies (where no control is applied).

\subsection{Multi-channel multi-frequency common pseudo-error equalizer}

To calculate the transfer functions of the common pseudo-error multi-frequency
algorithm, we have to proceed as in the single-frequency case, but now
addressing a system with $(K+J)\times L$ equations and $(K+J)\times L$ unknowns, based on the expressions of the $K$ signals at the error sensors and the $L$ signals generated by each of filter that controls each harmonic in each of the $J$ loudspeakers. These equations will be given by the $K$ following equations (one for each sensor):
\begin{equation}\begin{array}{r}
\displaystyle{E_k(z)=D_k(z)+\sum_{l=1}^L\sum_{j=1}^J(1-\gamma_{lj})Y_{lj}(z)C_{jk}(z)},
\label{Eftec}\end{array}
\end{equation}

where $Y_{lj}(z)$ is the Z-transform of the signal corresponding to the $l$-th harmonic generated by the filter that feeds the $j$-th loudspeaker, and derived from filtering the reference signal of the $l$-th frequency by the coefficients given by (\ref{Wlmspsecom}). Considering that the reference signal for
each frequency has unit amplitude and zero initial phase, we work
in the same way as the previous section for the single-frequency case, and thus obtain:
\begin{equation}
{Y}_{lj}(z)=\sum_{k=1}^K G_{ljk}(z)E'_k(z), \label{Yftec}
\end{equation}

where we will now have different auxiliary functions
particularized for each frequency $l$, and defined according to:
\begin{equation}
{G}_{ljk}(z)=\displaystyle{-2\mu_l\frac{\widetilde{A}_{ljk}(1-\gamma_{lj})[z \cos(\omega_{l}-\widetilde{\phi}_{ljk})-\cos(\widetilde{\phi}_{ljk})]}{(1-\beta_{lk})(z^{2}-2z \cos(\omega_{l})+1)}}.
\label{Gftec}
\end{equation}

Introducing the Z-transform of the pseudo-error signal defined in eq.(\ref{pseudodcom}) into eq.(\ref{Yftec}) and
grouping the terms that depend on ${Y}_{lj}(z)$, we can write:
\begin{small}
\begin{equation}\begin{array}{l}
\displaystyle{Y_{lj}(z)\left(1-\sum_{k=1}^K
\frac{1-\gamma_{lj}}{1-\beta_{lk}}\beta_{lk}G_{ljk}(z)\widetilde{C}_{jk}(z)\right)=}\\\displaystyle{\sum_{k=1}^K
G_{ljk}(z)\left\{E_k(z)+\sum_{q=1,q\neq l}^L \sum_{r=1,r\neq j}^J
\frac{1-\gamma_{rj}}{1-\beta_{qk}}\beta_{qk}\widetilde{C}_{jk}(z)Y_{qr}(z)\right\}.}\end{array}
\label{Yftec2}
\end{equation}
\end{small}

Thus the remaining $L\times J$ equations are obtained. These equations should be divided by $E_m(z)$ to calculate the $m$-th transfer function. Thus, with the aim to obtain the transfer function $H_1(z)$, we can suitably rearrange the data of the system of equations  (in similar way of the system of equations in (\ref{comp})) as:
 \begin{equation}
\begin{array}{l l l}
\left[ \begin{array}{c c } \mathbf{A}_{K\times K} &\mathbf{ B}_{K\times LJ} \\
\mathbf{D}_{LJ\times K} & \mathbf{E}_{LJ\times LJ}
 \end{array} \right] & \left[\begin{array}{c}
\mathbf{f}_{K\times 1} \\
\mathbf{g}_{LJ\times 1}
 \end{array}\right] &  =\left[\begin{array}{c}
\mathbf{u}_{K\times 1} \\
\mathbf{v}_{LJ\times 1}
 \end{array}\right],
 \end{array}
\label{compec}
\end{equation}

where the ${K\times K}$ elements of the matrix $\mathbf{A}_{K\times K}$ would be calculated according to (\ref{eleA}) as before. The elements of matrix $\mathbf{B}_{K\times LJ}$ can be written as an array of $L$ blocks of size $K\times J$, of the form:
$\mathbf{B}_{K\times LJ}=[\mathbf{B}1_{K\times
J}\;\mathbf{B}2_{K\times J}\;\ldots\;\mathbf{B}L_{K\times J}]$,
being the elements located in the row $f$ and column $c$ of each submatrix  $\mathbf{B}l_{K\times J}$ given by
\begin{equation}\label{eleBlec}
      bl_{fc}=(1-\gamma_{lc})C_{cf}(z).
\end{equation}

We consider again that the elements of the matrix $\mathbf{D}_{LJ\times K}$
can be written as a matrix of $L$ blocks of size $J\times K$, of the form $\mathbf{D}_{LJ\times
K}=[\mathbf{D}1^T_{J\times K}\;\mathbf{D}2^T_{J\times K}\;\ldots
\mathbf{D}L^T_{J\times K}]^T$, where the elements of each block
are given by
\begin{equation}\label{eleDlec}
\begin{array}{ll}
  dl_{fc}=0, &  c=1,  \\
    dl_{fc}=-G_{lfc}(z), &  c\neq 1.
\end{array}
\end{equation}

 Matrix $\mathbf{E}_{LJ\times LJ}$ can be rearranged as a matrix of $LJ\times L$ blocks of size
$J\times J$ each. Each of these blocks,
located in the row $F$ and column $C$, is denoted as $\mathbf{E}FC_{J\times J}$, and
its elements are defined as follows:
\begin{equation}\label{eleElec}
\begin{array}{ll}
\mathrm{if}\;F=C=l &\\
  \displaystyle{e_{fc}=1-(1-\gamma_{lc})\sum_{k=1}^K \frac{\beta_{lk}}{1-\beta_{lk}}G_{lfk}(z)\widetilde{C}_{ck}(z)}, &  f = c,  \\
   \displaystyle{e_{fc}=-(1-\gamma_{lc})\sum_{k=1}^K \frac{\beta_{lk}}{1-\beta_{lk}}G_{lfk}(z)\widetilde{C}_{ck}(z)},&  f\neq
   c, \\
\mathrm{if}\;F\neq C &\\
   \displaystyle{e_{fc}=-(1-\gamma_{Cc})\sum_{k=1}^K \frac{\beta_{Ck}}{1-\beta_{Ck}}G_{Ffk}(z)\widetilde{C}_{ck}(z)}.
\end{array}
\end{equation}

The vector of unknowns $\mathbf{f}_{K\times 1}$ will coincide with that described in (\ref{vF}), and $\mathbf{g}_{LJ\times 1}$ will be:
\begin{equation}
\mathbf{g}_{LJ\times
1}^T=\left[\frac{Y_{11}(z)}{E_1(z)}\;\frac{Y_{12}(z)}{E_1(z)}\;\ldots\;\frac{Y_{1J}(z)}{E_1(z)}\;\ldots\;\frac{Y_{LJ}(z)}{E_1(z)}\right].
\label{vglec}
\end{equation}

In the same way, the independent terms defined in
$\mathbf{u}_{K\times 1}$ will coincide with those already shown in
(\ref{eleU}). While the $LJ$ terms of $\mathbf{v}_{LJ\times 1}$ (rewriting this vector as a concatenation of $L$ vector of size $J\time 1$)
can be calculated for each element of each vector $\mathbf{v}l_{J\times 1}$ as
\begin{equation}\label{eleVlec}
\begin{array}{c}
  vl_{f}=G_{lf1}(z).
\end{array}
\end{equation}
Solving the proposed system for each frequency of interest,
we obtain the transfer function $H_1(z)$ as the inverse of the first element of $\mathbf{f}_{K\times 1}$. The rest of the transfer functions
would be obtained in a similar way by properly interchanging the
data of the matrices and vectors of
(\ref{compec}).

\subsection{Multi-channel multi-frequency multiple pseudo-error equalizer}

The computation of the transfer functions for the multiple pseudo-error strategy is carried out in the same way as described in the previous section, except that the signals generated by the adaptive filters are now given by:
\begin{equation}
{Y}_{lj}(z)=\sum_{k=1}^K G_{ljk}(z)E'_{lk}(z), \label{Yftem}
\end{equation}

where the Z-transform of the new pseudo-error signal is defined.
This modifies the approach of the $J\times L$ equations that
involve the functions ${Y}_{lj}(z)$, which is given now by:
\begin{small}
\begin{equation}\begin{array}{l}
\displaystyle{Y_{lj}(z)(1-\sum_{k=1}^K
\frac{1-\gamma_{lj}}{1-\beta_{lk}}\beta_{lk}G_{ljk}(z)\widetilde{C}_{jk}(z))=}\\\displaystyle{\sum_{k=1}^K
G_{ljk}(z)\left\{E_k(z)+\sum_{r=1,r\neq j}^J
\frac{1-\gamma_{rj}}{1-\beta_{lk}}\beta_{lk}\widetilde{C}_{jk}(z)Y_{lr}(z)\right\}.}\end{array}
\label{Yftem2}
\end{equation}
\end{small}

Thus, we can use all the functions, matrices and vectors defined in the previous section with the exception of the matrix $\mathbf{E}_{LJ\times LJ}$, which would be simplified as follows:
\begin{equation}\label{eleElem}
\begin{array}{ll}
\mathrm{if}\;F=C=l &\\
  \displaystyle{e_{fc}=1-(1-\gamma_{lc})\sum_{k=1}^K \frac{\beta_{lk}}{1-\beta_{lk}}G_{lfk}(z)\widetilde{C}_{ck}(z)}, &  f = c  \\
   \displaystyle{e_{fc}=-(1-\gamma_{lc})\sum_{k=1}^K \frac{\beta_{lk}}{1-\beta_{lk}}G_{lfk}(z)\widetilde{C}_{ck}(z)}, & f\neq
   c \\
\mathrm{if}\;F\neq C &\\
   e_{fc}=0.
\end{array}
\end{equation}
Therefore, matrix $\mathbf{E}_{LJ\times LJ}$ is a block diagonal matrix, whose only nonzero matrices are those of the blocks on the diagonal.

\section{Computation of the transfer functions' poles}
The transfer functions defined in (\ref{eq:Hft}) can be alternatively expressed as the following quotient of polynomial:
\begin{equation}
H_k(z)=\frac{N_k(z)}{M_k(z)}=\frac{N_k(z)}{(1-p_{k1} z^{-1})(1-p_{k2} z^{-1})\cdots (1-p_{k2L} z^{-1})},
\end{equation}
where it can be shown that the numerator and denominator have the same order (2L) and the poles appear as conjugate pairs.

The poles are inside the unit circle for stable systems, nevertheless their arguments and phases depend on: the ANE parameters, the estimation of the acoustic paths and the values of the control frequencies.

There exist a pair of conjugate poles associated to each control frequency for each transfer function, as it can be observed from the other sections analysis. The values of the poles clearly influence the ANE behaviour, mainly the duration of the transient stage. The poles' values can be numerically estimated by searching for the maximum of the transfer function modulus.

Although the poles' phases displace from the angle of the corresponding control frequency due to the interaction between the different poles and zeroes, the angles of the control frequencies are pretty good estimations of the poles' phases in most practical cases. Thus the estimation of the poles argument can be carried out by searching the position of the maximum values of the transfer functions following the radial lines corresponding to the control frequencies' angles.

\begin{figure}[ht]
\begin{center}
\begin{tabular}{cc }
\includegraphics[width=8cm]{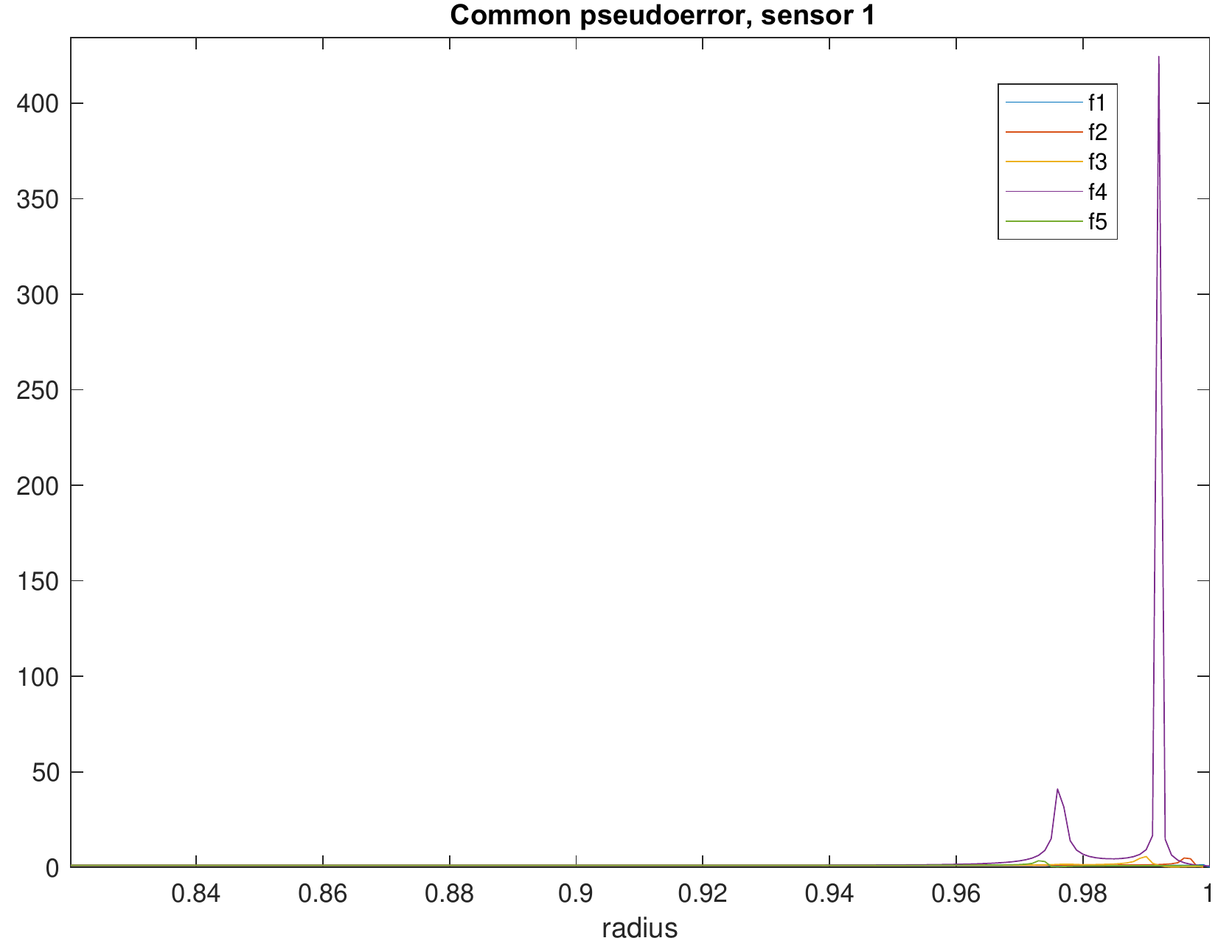} &
\includegraphics[width=8cm]{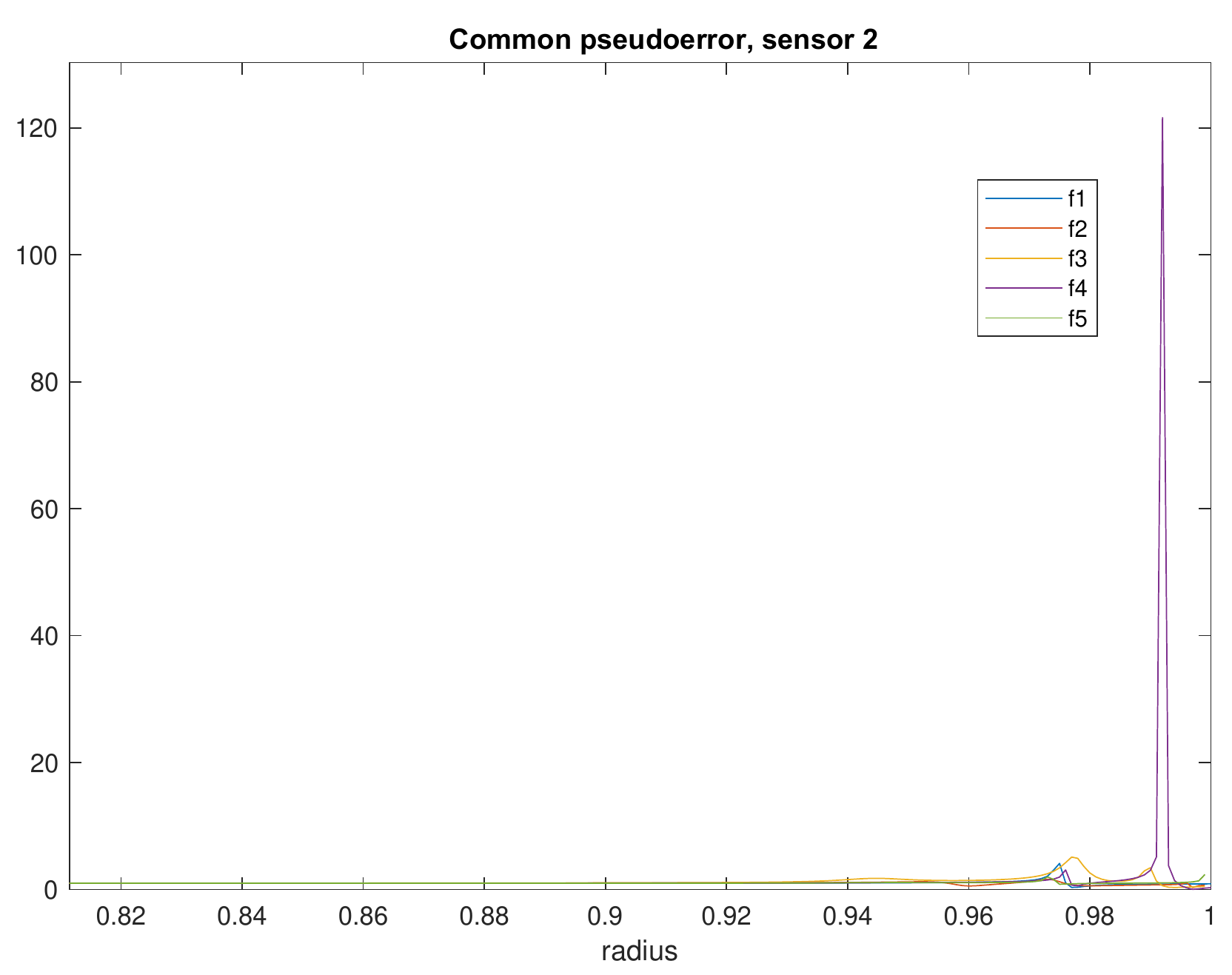}\\
\includegraphics[width=8cm]{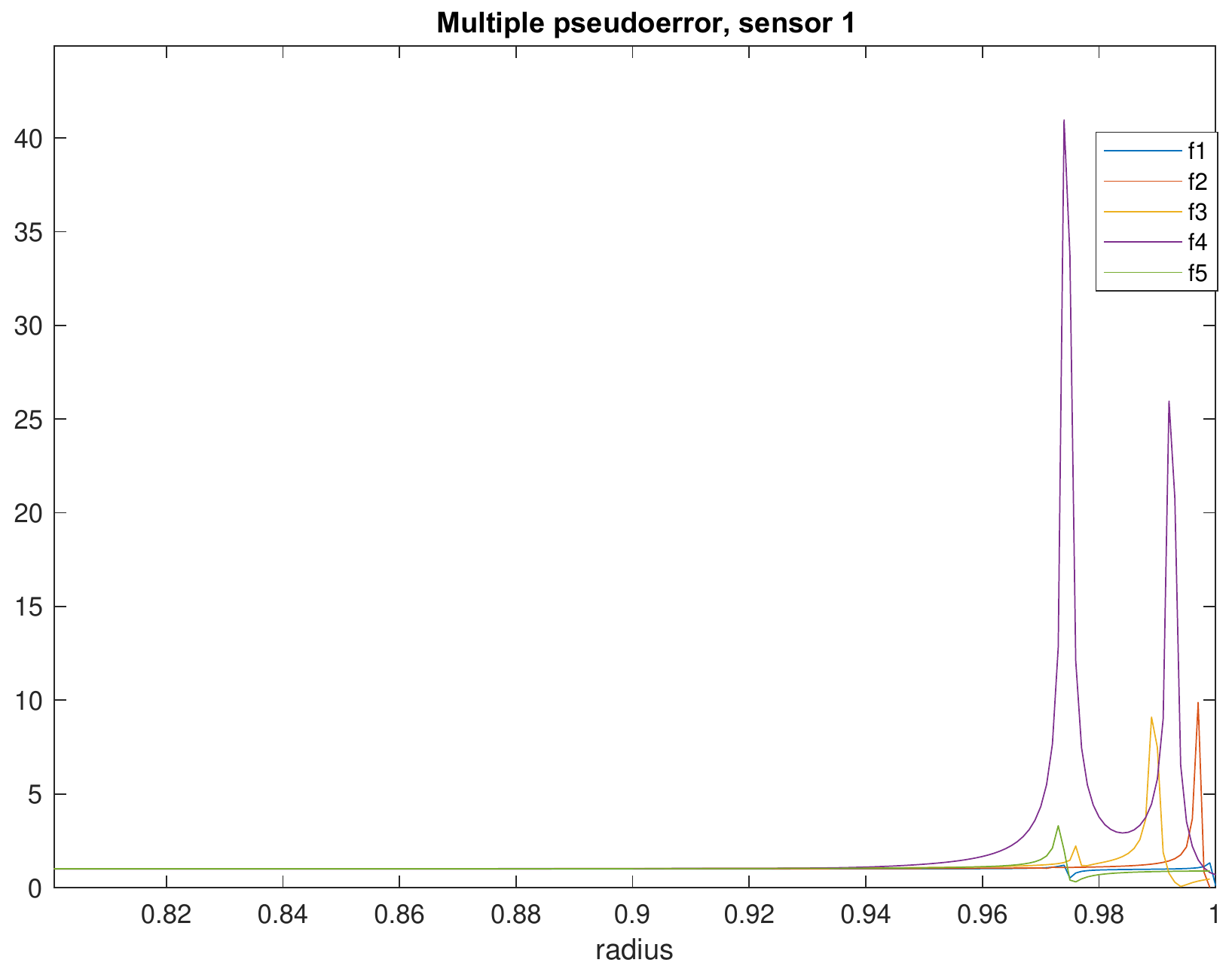} &
\includegraphics[width=8cm]{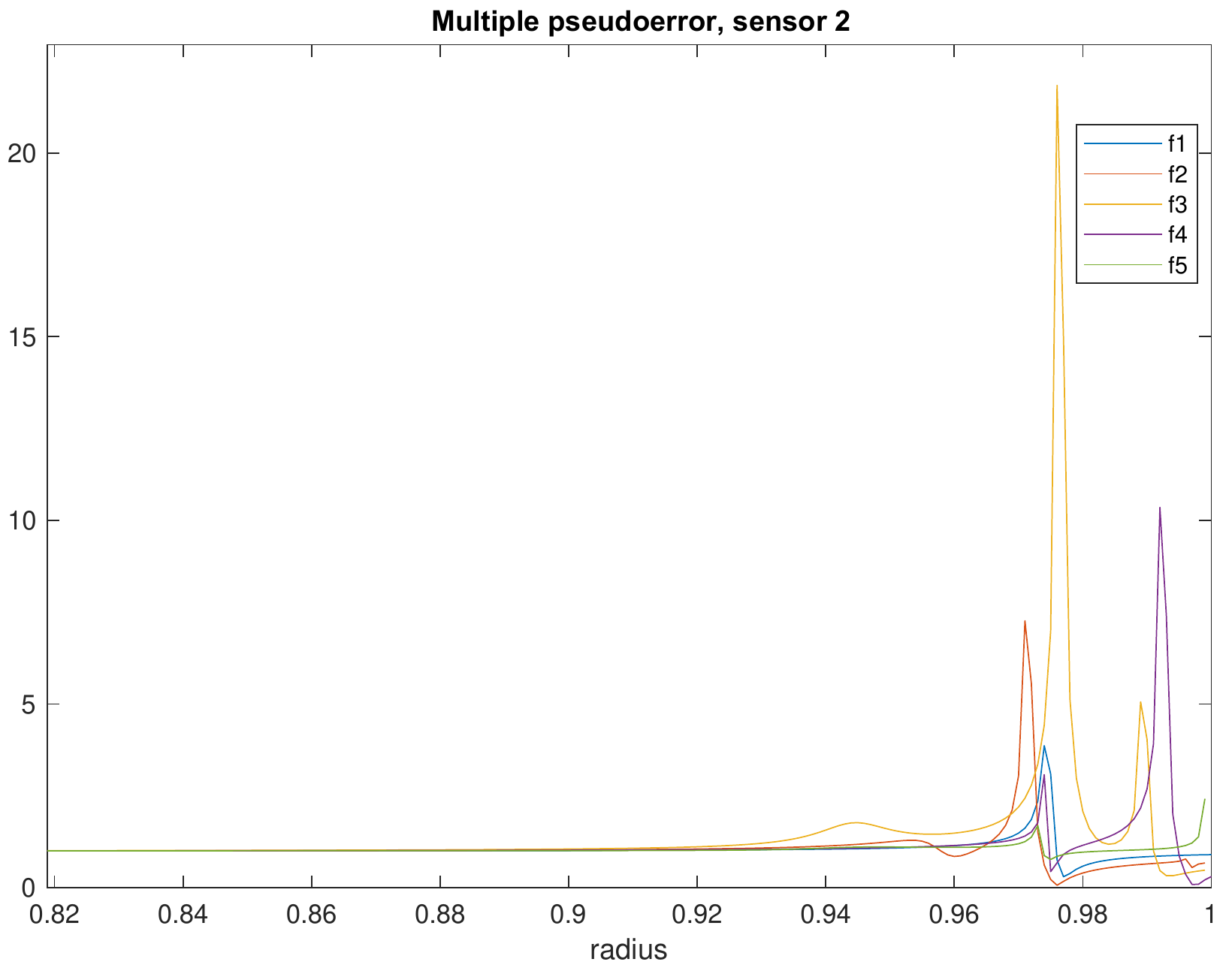}
\end{tabular}
\end{center}
\caption{Transfer functions' absolute values over a radial line at angles given by $f_l\in\left\{0.05,0.15,0.25,0.35,0.45\right\}$ for: a $2\times 2$ ANE, both pseudoerror strategies, and, equalization profiles given by: $\beta_1=[0.1, 0.3, 0.5, 0.7, 0.9]$ for sensor $1$, and, $\beta_2=[0.9, 0.7, 0.5, 0.3, 0.1]$ for sensor $2$.} \label{figmaxradius}
\end{figure}

Figure \ref{figmaxradius} shows the absolute values of the transfer functions of a $2\times 2$ ANE over the radial lines corresponding to the $5$ control frequencies $f_l\in\left\{0.05,0.15,0.25,0.35,0.45\right\}$, these are the values at $z=re^{j2\pi f_l}$ with $0<r<1$. The equalization profiles are given by: $\beta_1=[0.1, 0.3, 0.5, 0.7, 0.9]$ for sensor $1$, and, $\beta_2=[0.9, 0.7, 0.5, 0.3, 0.1]$ for sensor $2$. It is clearly illustrated by Figure~\ref{figmaxradius} that the transfer functions exhibit maximum values at given positions that allow to estimate the poles' arguments.

\section{Conclusions}
\label{s:conclusiones}

In this work we have extended the analysis of the multi-channel ANE for multi-frequency noise signals by mean of the calculation of its transfer functions. This equalizer allows different equalization gains for each sensor and for each frequency component of the noise, but it has a unrestrained behavior at the frequencies range out of the target of control, mostly depending on the their configuration parameters and the frequency response of the acoustical secondary paths. Moreover, in case of the frequencies under control are very close, the performance of the equalization could be adversely affected. To predict the behaviour of the system at the whole frequency range, the transfer function must be evaluated. Due to the acoustic coupling, the transfer functions only exhibit a closed expression for a single channel system. Thus, we have presented how to obtain these transfer functions for a generic multi-channel ANE. Firstly, we have depicted the methodology for a multi-channel single-frequency ANE and then we extend it for the multi-frequency case considering two previously proposed strategies (the common pseudo-error and the multiple pseudo-error). The numerical computation of the proposed method to calculate these transfer functions help to understand the behaviour of these equalizers, even at transient state through the estimation of the magnitude of the poles of the transfer functions.

\section*{Acknowledgments}
This work has been partially supported by EU together with Spanish Government and Generalitat Valenciana through RTI2018-098085-BC41 and PID2021-125736OB-I00 (MCIU/AEI/FEDER), RED2018-102668-T and PROMETEO/2019/109.

\bibliographystyle{unsrt}

\end{document}